\documentclass[journal]{IEEEtran}
\ifCLASSINFOpdf
\else
\fi
\usepackage[cmex10]{amsmath}
\begin{document}
%
\title{Raptor Codes and Cryptographic Issues}
%
%
%

\author{Mikko~Malinen,~\IEEEmembership{Student~Member,~IEEE.}}
\maketitle

\begin{abstract}
In this paper two cryptographic methods are introduced. In the first method the precence of a certain size subgroup of persons can be checked for an action to take place. For this we use fragments of Raptor codes delivered to the group members. In the other method a selection of a subset of objects can be made secret. Also, it can be proven afterwards, what the original selection was.
\end{abstract}


%
\IEEEpeerreviewmaketitle

\section{Checking the presence of a certain size subgroup of persons}
%
%
%
%
\IEEEPARstart{C}{onsider,} that we have a group of persons. For an action to take place, we require that at least $s$ members of the group are present. The action could be firing a weapon, or we could perform some action for which we need a subset of a certain size of board members of a company, or just some action for which to take place we need to check the presence of a subset of certain size of some interest group.\\
\ \\
This checking of presence could be made by Raptor codes (for Raptor codes, see $[$1$]$). We could deliver in advance fragments of Raptor code to each member of the group. The fragment size should be such that we obtain the needed key (which is decoded from concatenated fragments) only if at least $s$ group members are present. From the properties of Raptor codes it follows that we can recover the key with any subset of size $s$ of fragments.\\
\ \\
In this paragraph we calculate the maximum number of group members whose presence could be checked by this method. Let $s$ be the number of persons whose presece is needed for an action to take place, and let $k$ be the length of the key (number of input symbols). $s$ fragments of Raptor code should be $1.1\cdot k$ in total length to be decodable with very high probability. As we know from $[$2$]$, 0.05 overhead can make the code decodable. We require here 0.1 overhead (this is the amount which is a design choice of at least some companies for their Raptor codes $[$3$]$). Also, the total length of $s-1$ fragments of Raptor code must be less than $k$ in length for the code to be non-decodable. Let's say that $s-1$ fragments is $0.99\cdot k$ in length. With one output symbol, at most one input symbol can be recovered. This is why it is not possible to decode whole message with $s-1$ fragments or less. From these we get
$$\lfloor \frac{\text{total length of fragments}}{\text{length difference between {\it s} and {\it s-1} fragments}}\rfloor$$
$$= \lfloor \frac{1.1k}{1.1k-0.99k}\rfloor =10$$ 
which is the maximum number of group members whose presence could be made required for an action to take place.\\
\ \\
It should be noted that the group size is almost not limited from above. The group could be for example the population of a nation. The group members may carry the coded fragments by memory sticks and the key is stored on a computer where the memory sticks will be attached. The group members can be even geographically distributed and for example they attach their memory sticks to their computers which are connected to Internet by a secure connection.\\
\ \\
This scheme could be implemented also by passwords. Raptor code method is better when passwords can not be used for some reason, for example if their space requirement is too large due to the big size of the group.\\
\ \\
A drawback of this method is that if the key has to be changed, then all the code fragments have to be changed also.

\section{Private subset of objects}
Let's begin the description of this application with a case. In Finland, for example, in national Lotto the customers guess 7 numbers out of 39. Let's suppose that the Lotto customers want to keep their selection of numbers secret, but send their selection to the Lotto company. This can be done so that to each of 39 Lotto numbers an integer $i$ is assigned, which is the product of two big random prime numbers $j$ and $k$. Also, an integer $l$ 0..9 is assigned to $i$, which is formed as follows: If the customer has selected number $i$, $l$ is the mod 10 sum of the digits of $j$ and $k$. For example, if $j$ and $k$ are 327...3 and 615...7 then $l$ is 3+2+7+...+3+6+1+5+...+7 (mod 10). In case the Lotto customer has {\it not} selected number $i$ then $l$ is formed so that we take the $l$ calculated as above and add 1 to it and take mod 10. In this way the customer has $i$ and $l$ associated to each Lotto number 1..39 and he or she sends these to the Lotto company. Because of computational difficulty of factorization of composite numbers formed this way the Lotto company can't know $j$'s and $k$'s and so is not able to check if they match with $l$ to know if the corresponding Lotto number is selected by the customer. The idea of a product of large primes is from the well known RSA $[$4$]$ public key cryptography algorithm.\\
\ \\
The Lotto company may send the file sent by the customer back to the customer decrypted by the company's private key in RSA manner. This "decrypted" file may serve as a receipt from the company.\\
\ \\
When the drawing of Lotto numbers have been done, the winning customers send their $j$'s and $k$'s to Lotto company so that it can check if the customer has won.\\
\ \\
Another application very similar with the Lotto example is predicting the election results without telling a priori which candidates he or she predicts to win. The correctness of the prediction can be verified after the election result. This way we can avoid the effect of self-fulfilling prophecies because the prophecies will be kept secret. This method is applicable to almost all situations where a subset of a set has to be chosen and the selection has to be kept secret.

\section*{References}

\ \\
$[1]$ A. Shokrollahi, "Raptor Codes", \emph{IEEE Transactions on Information Theory}, Vol. 52, No. 6, 2006\\
\ \\
$[2]$ D. J. C. MacKay, "Fountain codes", \emph{IEEE Proceedings: Communications}, Vol. 152, No. 6, Dec. 2005, pp. 1062-1068\\
\ \\
$[3]$ S. Siikavirta, private conversation\\
\ \\
$[4]$ R. L. Rivest, A. Shamir, and L. Adleman, "A Method for Obtaining Digital Signatures and Public-Key Cryptosystems", \emph{Communications of the ACM}, Vol. 21, Feb. 1978, pp. 120-126\\

\end{document}